\documentclass[aps,showpacs,twocolumn,toolkits,nofootinbib]{revtex4}
\usepackage{mathrsfs}
\usepackage{amsmath}
\usepackage{graphicx}
\usepackage{color}
\usepackage{epsfig}
\usepackage{subfigure}

\begin{document}

\title{The Schwinger mechanism revisited}

\affiliation{Department of Physics, University of Maryland,
College Park, Maryland 20742-4111, USA}

\author{Thomas D. Cohen}
\email{cohen@physics.umd.edu}

\affiliation{Department of Physics, University of Maryland,
College Park, Maryland 20742-4111, USA}

\author{David A. McGady}
\email{dmcgady@umd.edu}

\affiliation{Department of Physics, University of Maryland,
College Park, Maryland 20742-4111, USA}


\begin{abstract}
The vacuum persistence probability, $P_{vac}(t)$, for a system of charged fermions in a fixed, external, and spatially homogeneous electric field, was derived long ago by Schwinger; $w \equiv - \log[P_{vac}(t)]/ V t$ has often been identified as the rate at which fermion-antifermion pairs are produced per unit volume due to the electric field.  In this paper, we separately compute exact expressions for both $w$ and for the rate of fermion-antifermion pair production per unit volume, $\Gamma$, and show that they differ.  While $w$ is given by the standard Schwinger mechanism result of  $w = \frac{(q E)^2}{4 \pi^3 } \, \sum_{n=1}^\infty  \frac {1}{n^2} \, \exp \left( -\frac{n \pi m^2 }{q E } \right )$, the pair production rate, $\Gamma$, is just the first term of that series.  Our calculation is done for a system with periodic boundary conditions in the $A_0=0$ gauge but the result should hold for any consistent set of boundary conditions.  We discuss, the physical reason why the rates $w$ and $\Gamma$ differ.
\end{abstract}

\maketitle
\section{Introduction}

Fermion-antifermion pair production from a static classical electric field, known as the Schwinger mechanism, has given rise to a vast literature, since its formulation in $1951$\cite{Schwinger}. Invoked to gain insights on topics as diverse as the string breaking rate in QCD\cite{CNN,Neuberger} and on black hole physics\cite{GR}, this mechanism has become a textbook topic in quantum field theory\cite{IZ}.  Topics such as back reaction\cite{BR} and finite size effects\cite{Wang_Wong} have been addressed. In a classic paper, Schwinger exactly calculated the rate at which the vacuum decays due to pair production in the external field. If the electric field is treated classically---\emph{i.e.} the formal limit of $q\rightarrow 0$, $E \rightarrow \infty$ with $q E$ fixed---the vacuum persistence probability is\cite{Schwinger}:
\begin{eqnarray}
P_{\rm vac}(t)& \equiv& |\langle {\rm vac}| U(t) |{\rm vac} \rangle |^2 = \exp(-w V t) \label{SF1}\\
{\rm with} \; \; w &=& \frac{(q E)^2}{4 \pi^3 } \, \sum_{n=1}^\infty  \frac {1}{n^2} \,
\exp \left( -\frac{n \pi m^2 }{q E } \right ), \label{SF2}
\end{eqnarray}
where $V$ is the spatial volume of the system and $w$ is the rate of vacuum decay per unit volume. 

While the Schwinger formula of Eq.~(\ref{SF2}) is very well known, the Schwinger mechanism has often not been fully appreciated in an essential way. The quantity $w$ in Eq.~(\ref{SF2}) appears to have a very natural interpretation as the rate of production of pairs per unit volume. Schwinger  suggested this interpretation in his original paper.  There is a plausible, if heuristic, argument in its favor\cite{CNN,IZ}.  Start by considering a more general situation in which the pair production rate can vary in space and time.  The vacuum persistence probability can then be written as
\begin{equation}
P_{\rm vac}=e^{ - \int {\rm d}^4x \ w(x) } \; .
\end{equation}
Next approximate the integral as a discrete sum over space-time cells of volume $\Delta v_i$ points centered at points $x_i$.  The vacuum persistence probability then becomes the limit of
\begin{equation}
P_{\rm vac}= \lim_{\Delta v \rightarrow 0} \prod_i e^{-w(x_i) \Delta v_i}=\prod_i (1- w(x_i) \Delta v_i) \; .
\end{equation}
Itzykson and Zuber\cite{IZ} note that this is the form expected from an independent contribution to the vacuum persistence coming from each cell if $w(x_i)$ is the local rate of pair production.  This is taken to confirm Schwinger's interpretation of $w$.  This argument {\it is} plausible; it is not surprising that a considerable body of literature has adopted Schwinger's interpretation and used $w$ to compute particle production rates.

However, as it happens, the interpretation is not correct: in general $w$ does {\it not}  give the rate of pair creation.  As a logical matter the rate characterizing the vacuum's decay (by the production of the first charged pair), is not necessarily the same as the continuous rate of pair production.  While the argument given above is highly plausible, it is also heuristic.  A more compelling approach is via a comparison of a direct computation of the pair production rate with $w$, the rate in Eq.~(\ref{SF2}).   Nikishov directly computed the pair production rate per unit volume long ago\cite{Niki} and found it to be
\begin{equation}
\Gamma = \frac{(q E)^2}{4 \pi^3 } \,
\exp \left( -\frac{\pi m^2 }{q E } \right ) \; ; \label{SF3}
\end{equation}
Remarkably it does {\it not} agree with $w$: the entire rate is given by the first term in the series for $w$.

Since $w$ is still commonly confused with the pair production rate, it is useful to rederive Nikishov's result in a physically transparent context which clearly illustrates why $\Gamma$ rather than $w$ is the rate of pair creation per unit volume.  This will first be done in the context of (1+1) dimensions where the issues are particularly clear; the generalization to (3+1) dimensions is straightforward.

Before embarking on the calculation, it is useful to discuss briefly why one might expect $\Gamma$ (the pair production rate per unit volume) to differ from $w$ (the rate characterizing the rate of vacuum decay per unit vacuum).  We begin with the trivial observation that pairs are created in distinct modes.  If the probability for any given mode to create a pair is constant in time (and thus uncorrelated with each other), one expects that total rate of pair creation to equal $w$. However, if there are temporal  correlations between likelihood for decay at various times for the various modes, then this need not be the case.

The precise time that a pair is created is ambiguous; however, once a pair is created and the fermion and antifermion become well separated, they each accelerate in the electric field.  Thus particles with large (mechanical) momentum are likely to have been created earlier than particles with small momentum. To the extent that the natural way to characterize modes is in momentum space, there are very large temporal correlations in pair creation.  As will be seen in detail in the explicit calculations, momentum space {\it is} the natural way to characterize modes for this problem and it is natural that $\Gamma$ differs from $w$. This makes clear why the argument of Ref.~\cite{IZ} that $w$ is the pair creation rate breaks down: there  it is assumed that the rate of pair production at distinct space-time points are independent.  Implicit in this is the notion that the pairs are created at well-defined points in space-time.  However, if the pairs are created at well-defined (or nearly well-defined) momenta they are delocalized in position.

\section{Pair creation in (1+1) dimensions}

Before embarking on the original problem, it is useful to consider the analogous problem in (1+1) dimensions.  The (1+1) dimensional case is somewhat simpler and the issues are more straightforward.  Moreover, the (1+1) dimension result serves as an important input into the (3+1) dimensional calculation.  One can adapt Schwinger's calculation\cite{Schwinger} for smaller dimensions\cite{1plus1, 2plus1}; in (1+1) dimensions $P_{\rm vac} (t)$ is
\begin{eqnarray}
P_{\rm vac}^{1+1}(t)& =&  \exp(-w^{1+1} L t) \nonumber \\
{\rm with} \; \; w^{1+1} &=& \frac{(q E)}{2 \pi } \, \sum_{n=1}^\infty  \frac {1}{n}\,
\exp \left( -\frac{n \pi m^2 }{q E } \right )\nonumber \\
& =& - \frac{(q E)}{2 \pi }\log \left(  1 - \exp \left( -\frac{ \pi m^2 }{q E }   \right )  \right )\label{SF1p1}
\end{eqnarray}
where $L$ is the length of the system.

The imposition of appropriate boundary conditions in both space and time is crucial in the treatment of this problem. Accordingly, it is critical to compute both $w$ and $\Gamma$, {\it with the same boundary conditions} so they can be directly compared with each-other.  We adopt the following strategy in choosing boundary conditions in space and time.

The most natural way to identify the number of pairs created is to formulate the problem in terms of an electric field which is switched on for a fixed period and then switched off.  One begins with the system in its vacuum state and the electric field turned off.  The electric field is turned on at $t=0$ and left on until $t=T$ and then switched off.  With the field now off, one can unambiguously use a free particle basis to count the number of pairs.  The act of turning the system on and off can yield transient effects; thus one needs to consider the limit of large $T$ ($T \gg m^{-1}$ , $T \gg qE^{-1/2}$) so that the steady state Schwinger mechanism dominates.

To simplify the computation it is useful to keep the electric field a constant over the entire system while making the system of finite length.  In this case the modes become discrete and are easily counted.  The large $L$ limit can be taken at the end.  Continuous acceleration of particles over extended time plays a key role in the analysis.  Thus it is important to choose spatial boundary conditions which allow a charged particle to continue accelerating as it reaches the end of the system. Dirichlet boundary conditions are thus inappropriate as a particle striking the end of the system will bounce back and, instead of continuing to accelerate, will then decelerate.  Continuous acceleration over long times can be achieved simply via the implementation of periodic boundary conditions.  However, periodic boundary conditions are sensible only when the Hamiltonian is invariant under translations by $L$. Because the gauge choice $A_0=0$ ensures translational invariance, we work in this gauge.  Of course, our results must be gauge invariant and we can work in any gauge.  However, if we choose to adopt a different gauge the boundary conditions needed to ensure continuous acceleration become time dependent.  Ultimately we can check whether this choice of boundary condition (and its associated gauge choice) is sensible by comparing the derived vacuum persistence probability with the Schwinger result of Eq.~(\ref{SF1p1}).
The Dirac equation becomes
\begin{align}
&\left ( \alpha (-i \partial_x -q A_x(t))  +  \beta m \right ) \psi(x,t)  =  i \partial_t \, \psi(x,t) \nonumber\\
 &A_x(t)=   \begin{cases} 0 & {\rm for} \; t<0 \\ -E t &  {\rm for}  \; 0 \le t<T\\  -ET & {\rm for} \;  T \le t \end{cases} \nonumber \\
 &\psi(x,t)  =  \psi(x+L,t)
\end{align}
where $\psi$ is a two-dimensional spinor and $\alpha$ and $\beta$ are the appropriate two-dimensional Dirac matrices.  There exists a complete set of a solutions of the form
\begin{align}
\psi(x,t) & =  e^{i p_k x} \chi_k(t) \nonumber \\
p_k & =  k \,  \frac{2 \pi}{L} \; \; \;({\rm integer} \; k) \nonumber\\
h_k(t)& =  \alpha(p_k -q A_x(t)) + \beta m  \nonumber \\
h_k(t) \chi_k(t) & = i \partial_t \, \chi_k(t)
\label{mode}\end{align}
where the restriction to integer $k$ comes from the periodic boundary conditions.  Finally, it is convenient to introduce a unitary transformation, $U_k(t)$, into a basis in which $h_k$ is diagonal
\begin{align}
  \chi_k(t) & = U_k(t) \Phi_k (t) \nonumber \\  \epsilon_k(t) & = \sqrt{(p_k + q A(t))^2+m^2} \nonumber \\
U_k^\dagger(t) h_k(t) U_k(t) &= \begin{pmatrix} \epsilon_k(t) & 0 \\ 0 & -\epsilon_k(t) \end{pmatrix}
\label{trans}\end{align}

For convenience we choose $T$ such that $T= j \tau$ with $\tau \equiv \frac{2 \pi}{qE L}$ and j a positive integer. It is easy to see that $\epsilon_{k}(T)=\epsilon_{k+j}(0)$: the spectrum of the system after $E$ is turned off is identical to the original spectrum, although the mode labels have changed. This is hardly surprising, as the system for $t \ge T$ is simply that of a free particle in the absence of an electric field---exactly as it is for $t < 0$. The restriction to integer $j$ ensures that the boundary conditions are the same.  At $t=0$ the system is in its vacuum state with all of its negative energy levels filled. Thus the appropriate boundary condition for the fermionic modes is $\Phi_k^T(0)= (0, 1)$.

The equation of motion for $\Phi_k^T(t) \equiv (\phi_k^+(t), \phi_k^-(t))$ is obtained from Eqs.~(\ref{mode}) and (\ref{trans}) and for $0 \le t \le T$ is  given by
\begin{align}
  \begin{pmatrix} \epsilon_k(t) &  \frac{q E m}{\epsilon_k^2(t)} \\  \frac{q E m}{\epsilon_k^2(t)} & -\epsilon_k(t) \end{pmatrix} \begin{pmatrix} \phi_k^+(t) \\ \phi_k^-(t) \end{pmatrix}& = i \partial_t  \begin{pmatrix} \phi_k^+(t) \\ \phi_k^-(t) \end{pmatrix} \nonumber \\
  {\rm with} \; \; \;  \begin{pmatrix} \phi_k^+(0) \\ \phi_k^-(0) \end{pmatrix} &= \begin{pmatrix} 0 \\ 1 \end{pmatrix} \; . \label{EOM}
\end{align}
By construction, $|\phi_k^+(T)|^2$ is the probability for the positive energy state of the $k^{\rm th}$ mode to be occupied after the field has been switched off; it represents the probability of pair creation for this mode. The expectation value for the total number of pairs produced, $N$, and the vacuum persistence probability are thus given by
\begin{align}
\langle N(T)\rangle &= \sum_k |\phi_k^+(T)|^2 \label{number}\\
P_{\rm vac}^{1+1}(T) &= \prod_k \left (1 - |\phi_k^+(T)|^2 \right )\nonumber \\
& =\exp \left (\sum_k \log\left (1 - |\phi_k^+(T)|^2 \right ) \right ) \; . \label{pvac}
\end{align}

To proceed further we need solutions of Eq.~(\ref{EOM}). An exact solution can be obtained in terms of parabolic cylinder functions with complex arguments. The form is cumbersome and will not be given here. The important point here is the asymptotic behavior valid when $|p_k| \gg m$, and $|p_k + qE T| \gg m$, in which case the solution reduces to
\begin{equation}
\begin{split}
|\phi_k^+(T)|^2 & = \theta (-p_k)\theta (p_k + qE T ) \exp \left( -\frac{ \pi m^2 }{q E } \right ) \\ & \times \left( 1 + {\cal O}\left(\frac{m}{|p_k|} ,\frac{m}{|p_k + qE T|}\right ) \right ) \; . \label{lz}
\end{split}
\end{equation}

The form of Eq.~(\ref{lz}) is not surprising. The equation of motion for the two level system in Eq.~(\ref{mode}) is precisely of the form considered long ago by Landau and Zener\cite{LZ}; $\exp \left( -\frac{ \pi m^2 }{q E } \right )$ is simply the  Landau-Zener transition probability which is valid when the initial and final levels are well separated on the scale of $m$.  Provided that $qE T \gg m$ the correction to Eq.~(\ref{lz}) is small except for a small number of modes.  Inserting this form into Eqs.~(\ref{number}) and (\ref{pvac}) and using the identity $\log(1 - \theta(y) \theta(z) x) = \theta(y) \theta(z) \log(1-x)$
yields
\begin{align}
\langle N(T)\rangle &=   \sum_k \theta(-p_k) \theta (p_k + qE T ) e^{-\frac{ \pi m^2 }{q E }}  \nonumber \\
&\times \left ( 1 + {\cal O} \left ( \frac{m}{qE T} \right ) \right ) \label{number2}\\
\log \left (P_{\rm vac}^{1+1}(T) \right )
& =\sum_k \theta(-p_k) \theta (p_k + qE T ) \log\left (1 -  e^{ -\frac{ \pi m^2 }{q E } } \right )\nonumber \\
& \times \left (1 + {\cal O} \left ( \frac{m}{qE T}\right )  \right ) \,  \; . \label{pvac2}
\end{align}
where the scale of  the correction term reflects the fraction of modes for which the correction to the leading term in Eq.~(\ref{lz}) is non-negligible.  From the definition of $p_k$ in Eq.~(\ref{mode}) one sees that $\sum_k \theta(-p_k) \theta (p_k + qE T )= \frac{qE T L}{2 \pi}$. Using this fact, and taking the long time limit to remove transient effects, we see that $P_{\rm vac}^{1+1}(t)$ in Eq.~(\ref{pvac2}) exactly reproduces the standard result for the vacuum persistence probability in Eq.~(\ref{SF1p1}), indicating that our boundary conditions were chosen consistently.

Further, the rate of pairs produced by the electric field in the long time limit is,
\begin{equation}
\Gamma^{1+1} \equiv \frac{\langle N \rangle}{L T} = \frac{q E}{2 \pi} \exp \left( -\frac {\pi m^2}{q E} \right ). \label{number3}
\end{equation}
Critically, we see that $\Gamma^{1+1}$, the rate of pairs produced per unit length per unit time, is {\it not} $w^{1+1}$ but rather the first term in the series as expected from Ref.~\cite{Niki}.

\subsection{The massless limit}

The $m \rightarrow 0$ limit in (1+1) dimensions illuminates the essential issues quite clearly.  From Eq.~(\ref{SF1p1}) it is clear that $w^{1+1}$ diverges as  $m \rightarrow 0$.  Thus, the vacuum persistence probability becomes zero.  If the interpretation of $w$ as the pair production rate were correct this would mean that the rate of pair production per unit length would, perversely, be infinite for massless particles.  However, the particle production rate per unit length is  given by $\Gamma^{1+1}_{m \rightarrow 0} = \frac {q E}{2 \pi}$ and is finite.

\begin{figure}
\centering
\includegraphics[width=3.0in]{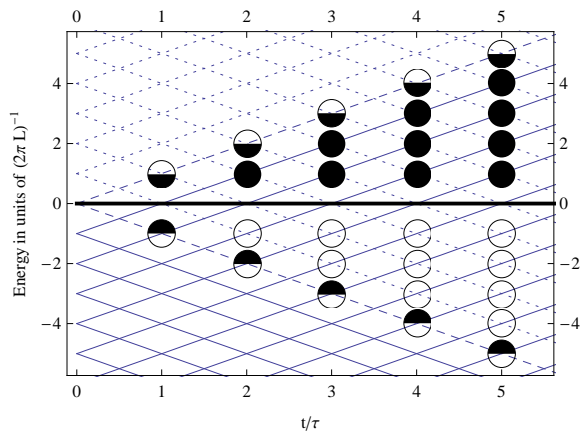}
\caption{Energy levels for massless (1+1) charged fermions in a constant electric field as a function of time in units of $\tau \equiv \frac{2 \pi}{qE L}$.  The dotted lines correspond to empty--initially positive energy--levels and the solid lines to filled--initially negative energy--levels; the dashed lines are the half-filled levels of the zero-mode.  The solid circles correspond to particles and the open circles to antiparticles; the half-filled circles correspond to the creation of a particle (antiparticle) with 50\% probability  }
\label{zeromass}
\end{figure}

It is easy to visualize this physically. With the boundary conditions used here, the energies of the modes for $m=0$ are given by $\pm (p_k + qE t)$ as shown in Fig.~\ref{zeromass}. Moreover the occupation numbers are preserved by the equation of motion; only the mass term induces transitions. There is a subtlety when $p_k+ qE t=0$; at which point the positive and negative energies cross. Since $\phi^+$ indicates the amplitude for the positive energy solution, the labels $\phi^+$ and $\phi^-$ switch at $p_k+ qE t=0$.  (Note that when any finite mass is put in, the level crossings are avoided.  However, as $m \rightarrow 0$ the Landau-Zener probability which the occupation number switches from the negative to positive level goes to unity and one reproduces the zero mass result.)

Initially, the system is in the vacuum state with all negative energy levels filled. Thus except for the special case of $k=0$, ~ $|\phi^+(t)|^2$ is given by $\theta(-p_k) \theta(qEt + p_k)$, exactly as one expects from Eq.~(\ref{lz}). The $k=0$ modes are special in that they have exactly zero energy at $t=0$; we take them to be half occupied. Consider the occupation of positive energy levels as time increases. As seen in Fig.~\ref{zeromass} when $t$ increases by $\tau \equiv \frac{2 \pi}{qE L}$ exactly one new level crosses from negative to positive and this corresponds to the creation of a pair. At $t=j \tau$ one has created on average $j-1/2$ pairs (the $1/2$ coming form the $1/2$ filled $k=0$ mode): the number of pairs increases linearly with time at a rate of $1/\tau = \frac{qE L}{2 \pi}$, precisely the rate in Eq.~(\ref{number3}). However, for $j \ge 2$ the vacuum persistence probability is strictly zero; at $j=2$ we {\it know} with unit probability that a pair has been created in the $k=-1$ mode.  This cleanly illustrates both the distinction between $\Gamma$ and $w$ and the critical role played by temporal correlations between modes.  It is the fact that pairs are created in different modes at different times which accounts for the difference.

A natural way to visualize the massless case in (1+1) dimensions is  to think of the (1+1) dimensional ``universe'' as a circular loop of radius $R=L/(2 \pi)$.  Suppose further that this loop is imbedded in a (3+1) dimensional world with a time-varying magnetic flux perpendicular to the loop and threading through its center.  This flux is turned on at $t = 0$, after which it linearly increases in time, for a time $t = j \tau$ after which it is held fixed; $\tau$ is defined as the amount of time for one flux quantum to thread through.  Because $\frac {\partial B}{\partial t} = - \nabla \times E$, this is identical to an $E$ field pointing tangent to the loop at all points, for a total time $\Delta t = j \tau$. The ground state of this system at $t=0$ is simply a filled Dirac sea of massless, charged particles---in just the same manner as the ground state of a normal wire represents a filled {\it Fermi} sea of massive, charged, particles (electrons). This system clearly has periodic boundary conditions, and thus gives another physical picture for understanding pair creation of charged particles in (1+1) dimensions.  While the $E-$field is on, magnetic flux-quanta thread through the ring at a rate of $1/\tau$.  Since the system is circular, the natural variable labeling modes is the angular momentum, $L_k$ (where $L_k$ is related to the momentum in the previous formulation via $L_k = R p_k$).  After $j$ flux quanta have thread through the loop, then $L_{k} \rightarrow L_{k+j}$. This both drives $j$ empty levels into the Dirac sea and $j$ filled levels out of the Dirac sea, creating $j$ fermion-antifermion pairs. The rate is $1/\tau $ and is precisely that given in Eq.~(\ref{number3}).

\subsection{A card game}

The distinction between the rate per length at which pairs over time, $\Gamma^{1+1}$, and the rate per length of vacuum decay, $w^{1+1}$, is essentially one of classical probabilities; for this (1+1)-dimensional system this can be simply illustrated in the context of a card game.

One might imagine that a casino has introduced a new game called ``vacuum.''  The rules of the game are simple: A very large number of decks of cards are shuffled together. One class of card is specified---this class could be rather general ({\it e.g.,~} all black cards or all diamonds), somewhat more specific ({\it e.g.,~} all sevens), or quite specific ({\it e.g.,~} aces of spades).  The pile of cards is said to be in the vacuum.  Cards are then turned over at a fixed rate of $1/\tau$.  The pile remains in the vacuum until the first card of the specified class is turned over.  Wagers are placed on how long the pile is in the vacuum and on how many cards of the specified class are turned over in a specified time.

It is easy to see that if one continues to turn over cards at the rate of $1/\tau$, the average number of cards in our specified class which are turned over after $t= j \tau$ (where $j$ is a positive integer) is
$\langle N_{\rm class} \rangle = f t/\tau$ where $f$ is the fraction of the cards in the class ({\it i.e.,} $f=1/2$ if the class is black cards; $f=1/52$ if the class consists of aces of spades).   Thus, $\gamma$, the average rate of production for cards in the class, is given by
\begin{equation}
\gamma = \frac{1}{\tau} f \; . \label{gamma}
\end{equation}
The probability of the pile remaining in the vacuum after $t=j \tau$ is obviously given by $P_{\rm vac}= (1-f)^{j}=(1-f)^{t/\tau}$.  This can be rewritten as
\begin{align}
P_{\rm vac}& = \exp (-\omega t) \nonumber \\
\omega & = - \frac{1}{\tau} \log (1 - f) = \frac{1}{\tau} \sum_n \frac{f^n}{n} \label{omega} \, .
\end{align}
It is obvious that the structure of Eqs.~(\ref{gamma}) and (\ref{omega}) are analogous to Eqs.~(\ref{number3}) and (\ref{SF1p1}) with $f$ playing the role of $e^{-\frac{\pi m^2}{q E}}$.

In the context of the game it is obvious why $\gamma$ and $\omega$ differ.  Suppose as an extreme case one considers as the class {\it all} cards.  The rate at which cards in the class are produced is obviously $1/\tau$; after the first card has been turned over the probability that the pile is in the vacuum is clearly zero.  This is the analog of the $m \rightarrow 0$ limit.

\section{Pair creation in (3+1) dimensions}

Having illustrated the essential distinction between the vacuum decay rate, $w$, and the rate of particle production, $\Gamma$, in the (1+1) dimensional system, we turn to the case of (3+1) dimensions. For simplicity we specify $\overrightarrow{E} = |E|\hat{z}$, and require solutions of the Dirac equation to satisfy periodic boundary conditions along the transverse direction (making the spectrum discrete and counting explicit).  In (3+1) dimensions there are two spin states for the fermions. However, in the energy basis analogous to Eq.~(\ref{trans}), the system is diagonal in spin; thus its sole effect is an overall factor of $2$ in $w$ and $\Gamma$.  In direct analogy to Eq.~(\ref{EOM}), the equation of motion for the mode specified by the spin state, $s$, and $\overrightarrow{k} = (k_x, k_y, k_z) = (n_x \frac {2\pi}{L_x}, n_y \frac {2\pi}{L_y}, n_z \frac {2\pi}{L_z})$ is
\begin{align}
  \begin{pmatrix} \epsilon_{\overrightarrow{k}}(t) &  \frac{q E m}{\epsilon_{\overrightarrow{k}}^2(t)} \\  \frac{q E m}{\epsilon_{\overrightarrow{k}}^2(t)} & -\epsilon_{\overrightarrow{k}}(t) \end{pmatrix} \begin{pmatrix} \phi_{\overrightarrow{k}, s}^+(t) \\ \phi_{\overrightarrow{k}, s}^-(t) \end{pmatrix}& = i \partial_t  \begin{pmatrix} \phi_{\overrightarrow{k}, s}^+(t) \\ \phi_{\overrightarrow{k}, s}^-(t) \end{pmatrix}\ \label{EOM3plus1}
\end{align}
with  $\epsilon_{\overrightarrow{k}}(t) = \sqrt{(k_z + qA_z(t))^2+(p_x^2+p_y^2+m^2)}$ the instantaneous energy of the mode $\overrightarrow{k}$ at time $t$.  As in (1+1) dimensions, the initial conditions are
$(\phi_{\overrightarrow{k}, s}^+(0), \phi_{\overrightarrow{k}, s}^-(0)) = ( 0 , 1 )$.  Note that since $A_{\mu}$ (in this gauge) does not change the transverse momentum, it acts exactly like an additional mass-term in the Landau-Zener tunneling rate. Thus, for convenience, we will define $m_T^2 \equiv m^2 + \overrightarrow{k}_T^2$.

All of the previous arguments leading up to Eq.~(\ref{lz}) go through, after the substitution $m^2 \rightarrow m_T^2$. Thus, we derive
\begin{eqnarray}
\frac {\langle N(T)\rangle}{VT} &=& 2 \sum_{k_T, k_z} \frac {\theta(-k_z) \theta (k_z + qE T )}{L_x L_y L_z T} e^{\left [ -\frac {\pi (m^2+ k_T^2)}{q E} \right ]} \label{number4}\nonumber\\
&=& 2 \frac{qE}{2 \pi} \frac{L_z T}{L_x L_y L_z T} \sum_{k_x,k_y} e^{ -\frac {\pi (m^2+k_x^2+k_y^2)}{qE}} \nonumber\\
&=& \frac {qE}{4 \pi^3} e^{ -\frac {\pi m^2}{qE}} \frac{L_x L_y}{L_x L_y} \int dk_x dk_y e^{ -\frac {\pi (k_x^2+k_y^2)}{qE} } \ .
\end{eqnarray}
Evaluation of this double integral exactly reproduces the rate of pair production given in Eq.~(\ref{SF3}). A similar calculation of log$[P_{vac}(T)] \equiv -w VT$  yields,
\begin{align}
&\log [P_{vac}(T)] \equiv -w VT ,\nonumber \\
&= 2 \, \sum_{k_z,k_T} \theta(-k_z) \theta (k_z + q E T ) \log [1 -  e^{-\frac {\pi (m^2+k_T^2)}{q E} }] \nonumber \\
&= \frac {q E L_z T}{\pi} \frac{L_x}{2 \pi}\frac{L_y}{2 \pi} \int dk_x dk_y \log [1- e^{ -\frac {\pi (m^2+k_T^2)}{q E} } ] \nonumber\\
&= - VT \frac {(qE)^2}{4 \pi^3} \sum_{n =1}^{\infty} \frac {1}{n} \int dk_x dk_y e^{ -\frac {n \pi (m^2+k_T^2)}{q E} } \ .
\end{align}
Evaluation of the double integrals, and identification of this quantity with $-w VT$, yields the expression for $w$ in Eq.~(\ref{SF2}). Exponentiating this rate, we derive the standard vacuum persistence probability in Eq.~(\ref{SF1}).

\section{Discussion}

In summary, we have explicitly shown that the rate associated with vacuum decay, $w$, and the rate of pair creation, $\Gamma$, differ: $\Gamma$ is given by the first term in the series for $w$.  This result is in a very real sense quite well known: it was first derived by Nikishov \cite{Niki} quite long ago.  It can be derived in various other elegant formulations\cite{Rafelski, NewRefs}. However, it is not as widely appreciated in the community as it should be: much of the literature in the field is still  based on Schwinger's initial assumption that $w$ gives the rate of pair production per unit volume.  It is hoped that the physically transparent way that this paper shows that $\Gamma$ differs from $w$ will help clarify the issue.  The massless case in (1+1) dimensions is particularly illuminating.  It is clear from Fig.~\ref{zeromass} why the pair production rate remains finite even while the vacuum persistence probability goes to zero at finite times indicating an infinite  value for $w$.   While this calculation was with periodic boundary conditions in the $A_0=0$ gauge, we expect the result to hold generically for any consistent set of boundary conditions (\emph{i.e.,} those that reproduce the Schwinger result for $w$). This will be investigated in future work.

In practice, the distinction between $w$ and $\Gamma$ is exponentially small for weak fields. However, for strong fields, $w$ exceeds $\Gamma$ by a factor as large as $\zeta(2) = \frac {\pi^2}{6} \sim 1.64$.  More generally, in $d$ space-time dimensions $w^d$ will exceed $\Gamma^d$ by a factor of $\zeta(d/2)$ in the strong field limit.

T.D.C.\ was supported by the United States D.O.E.\ through grant number DE-FGO2-93ER-40762.  We thank S. Nussinov and I. Shovkovy for useful discussions, and C. Goebel for a careful reading of the manuscript. J.~Rafelski was very helpful in pointing out the relevant literature.

\end{document}